\documentclass[a4paper]{article}
\usepackage{amstex}
\usepackage{array}

\newcommand{\beq}{\begin{equation}}
\newcommand{\eeq}{\end{equation}}
\newcommand{\bea}{\begin{eqnarray}}
\newcommand{\eea}{\end{eqnarray}}

\renewcommand{\theequation}{\thesection.\arabic{equation}}

\newcounter{subequation}[equation]
\makeatletter
\expandafter\let\expandafter\reset@font\csname reset@font\endcsname
 
\def\subeqnarray{\arraycolsep1pt
    \def\@eqnnum\stepcounter##1{\stepcounter{subequation}
        {\reset@font\rm(\theequation\alph{subequation})}}\eqnarray}

\makeatother
 
\begin{document}
\thispagestyle{empty}
\parskip=12pt
\raggedbottom

\renewcommand{\thefootnote}{\fnsymbol{footnote}}
\def\mytoday#1{{ } \ifcase\month \or
 January\or February\or March\or April\or May\or June\or
 July\or August\or September\or October\or November\or December\fi
\space\number\day ,
 \space\number\year 
;\space time: \number\time
}

\noindent
\hspace*{9cm} BUTP/97--13\\
\vspace*{1cm}
\begin{center}
{\LARGE Fixed-Point Actions in 1-Loop\\[3mm]
 Perturbation Theory}%
\footnote{Work supported in part by Schweizerischer Nationalfonds\\ 
and by Iberdrola, Ciencia y Tecnologia, Espana}

\vspace{1cm}

Peter Hasenfratz and 
Ferenc Niedermayer\footnote{On leave from the Institute of Theoretical
Physics, E\"otv\"os University, Budapest}
\\
\vskip 1ex
Institute for Theoretical Physics \\
University of Bern \\
Sidlerstrasse 5, CH-3012 Bern, Switzerland \\
\vspace{0.5cm}

May 1997 \\ \vspace*{0.5cm}

\nopagebreak[4]

\begin{abstract}
\noindent It has been pointed out in recent papers that the example considered
earlier in the $O(N) ~\sigma$-model to test whether fixed-point actions
are 1-loop perfect actually checked classical perfection only. To clarify
the issue we constructed the renormalized trajectory explicitly in 1-loop
perturbation theory. We found that the fixed-point action is not exactly 1-loop
perfect. The cut-off effects are, however, strongly reduced also on 
the 1-loop level relative to those of the standard and tree level improved
Symanzik actions. Some points on off- and on-shell 
improvement, Symanzik's program and fixed-point actions are also discussed.
\end{abstract}

\end{center}
\eject
\renewcommand{\thefootnote}{\arabic{footnote}}
\setcounter{footnote}{0}

 \section{Introduction}
\setcounter{equation}{0}

The fixed-point (FP) action, which lies at the beginning of the renormalized
trajectory (RT) in an asymptotically free theory, defines a classically
perfect regularization \cite{HN1}: its classical solutions (instantons) are
scale invariant and in quadratic approximation in the fields the spectrum is
exact. It has been demonstrated in different models that the cut-off efffects
are strongly reduced also in the quantum theory 
\cite{HN1,DHHN2,BWIE,AP,RB,DHZ}.

The renormalization group (RG) transformation linearized around the FP (tree
level) has a marginal direction. One can raise some formal RG arguments
\cite{KW,DHHN1} for the statement that on the 1-loop level the FP action 
begins to move
along this marginal direction without changing its form. That would imply that
the FP action is automatically 1-loop perfect. The formal RG arguments are not
really convincing, however. For this reason an explicit example has been
constructed in ref.~\cite{FHNP}: it has been shown that the finite volume mass
gap $m(L)$ in the $d=2$ non-linear $\sigma$-model is free of cut-off effects
when calculated with the FP action in 1-loop perturbation theory. 
The mass gap is, however, a special quantity: any action gives cut-off
independent results for $m(L)L$ on the tree level. It has been conjectured and
illustrated through examples recently \cite{RV,CP} that in this case tree
level improvement will become effective on the 1-loop level cancelling the
cut-off effects there.
The mass gap therefore is not appropriate to test the issue at hand.

The mass gap $m(L)$ is determined by the large euclidean time decay of the zero
momentum ($p=0$) two-point function. On the tree level this leads to an
effective one-dimensional propagator which, for any action, has no power like
cut-off corrections. This is a special situation which does not remain true at
$p \neq 0$. The smallest energy $E(L,p)$ in the $p \neq 0$ channel defined by
\begin{equation}
C(\tau;p)=\frac{1}{L^2}\sum_{x^1,y^1}{\rm e}^{ip(x^1-y^1)}
\langle {\bf S}(x) {\bf S}(y) \rangle_{x^0=-y^0=\tau}
\sim B \exp\{-E(L,p) 2\tau\}
\label{CC}
\end{equation}
behaves as usual: for a generic action cut-off effects occur already on the
tree level and additional cut-off corrections are generated by perturbation
theory. We investigate here $E(L,p)$ with $p={2 \pi}/L$ for  different FP
actions, and for the tree level improved Symanzik and  standard actions. 
We found small cut-off
effects in the FP action predictions. In order to distinguish these effects
clearly from the systematic errors entering in the construction of the FP
action itself, we studied the RT (quantum perfect action) in 1-loop
perturbation theory. We identified explicitly the difference between the FP
action and the quantum perfect action and have shown that this difference is
responsible for the small cut-off effects seen.

In most of the applications of the FP action until now such block 
transformations were used where the average of the fine variables in 
a block was allowed to fluctuate around the block variable. 
This fluctuation is governed by a parameter $\kappa$ which was optimized 
to obtain a short range FP action. As we shall discuss in Section 4, 
in the formal RG considerations in ref.~\cite{DHHN1} the $\kappa=\infty$ 
case is special. 
Unfortunately, it is difficult to test this case because, for most of the
block transformations, the FP is not sufficiently short ranged at
$\kappa=\infty$, which increases the systematic errors of the
calculation. We studied the FP action proposed in \cite{HB} which has a short
ranged quadratic part at $\kappa=\infty$. We found, however, that the quartic
part and the quantum corrections to the action are much less compact than for
the optimised transformation considered in ref.~\cite{HN1}.
This might be related to the fact that
the block transformation in ref.~\cite{HB} is close to a decimation, and might
explain the poor numerical results obtained there.
We found strong indications that this action also produces
cut-off effects in 1-loop perturbation theory, but, due to the more extended
range of the action, the systematic
numerical errors were larger than before.  
In order to reach a definite conclusion also for the $\kappa=\infty$ case, 
we considered another physical quantity: the free-energy density as 
a function of the chemical potential. This calculation can
be done almost completely analytically, no numerical systematic errors
influence the results. On the other hand, introducing the chemical potential
raises delicate theoretical questions which we were not able to clarify
completely. We think, however, that the results, which show small
cut-off effects on the 1-loop level in this case as well, are correct.

We recapitulate and extend the arguments on the mass gap \cite{RV,CP} in
Section 2. We use this occasion to clarify some issues on off-mass shell
versus on-mass shell improvement and on the relation between Symanzik
improvement and the FP action. The energy $E(L,p)$ in 1-loop perturbation
theory is discussed in Section 3. In Section 4 we consider the relation
between the FP action and the marginal operator which makes the
$\kappa=\infty$ case special. In Section 5 we consider the steps leading to
the RT in 1-loop perturbation theory. Section 6 treats the dependence 
of the free energy density on the chemical potential. The numerical results 
are collected in Section 7.

Our main conclusion was already mentioned before: the FP action is not 1-loop 
quantum perfect. The authors express regret for having made incorrect statements on
this point earlier.
On the other hand, as expected intuitively, the FP action,
which is classically perfect, generates small cut-off effects on the 1-loop
level compared to
those of the standard nearest neighbour action. It is interesting to remark in
this context that the standardly used next-to-nearest neighbour realization of
the tree level improved Symanzik action gives a factor of $\sim 2$ 
larger $O(a^2)$ cutoff effect in $E(L,p)$ on the 1-loop level than the unimproved 
nearest neighbour action.

Some questions remained open. We are not able to identify where the formal RG
arguments go wrong. In ref.~\cite{KW} no details are given beyond eq.~(20) 
and the short paragraph following it. In the arguments of ref.~\cite{DHHN1} 
there are several questionable points. It is assumed that the FP action 
and the marginal operator on the classical level are identical. 
As discussed in Section 4 this is true only at $\kappa=\infty$. 
We find, however, even in this case cut-off effects. The formal arguments 
in \cite{DHHN1} rely further on the assumption that the eigenoperators 
of the linearized RG transformation around the FP form a complete system. 
This is not true in general \cite{KWJK}. Finally, it is not quite clear 
what the condition `close to the FP' used in the discussion means when 
the coupling constant in front of the action takes the value infinity at
the FP.

\section{The mass gap to 1-loop order}
\setcounter{equation}{0}

Consider the finite-volume mass gap $m(L)$ defined on a strip $0\le x \le L$,
$-\infty < t < +\infty$, with periodic boundary condition in $x$.  In
perturbation theory this has an expansion \cite{L,LWW,RV}
\begin{equation}
m(L)L=\frac{N-1}{2} g^2 \sum_{l=0}^{\infty} A_l(L) g^{2l}
\label{mPT}
\end{equation}
with
\begin{equation}
A_l(L)=\sum_{n=0}^\infty \left( \frac{a^2}{L^2}\right)^n
\sum_{p=0}^l a_{lp}^{(n)} \left( \ln \frac{L}{a}\right)^{l-p}
\label{Al}
\end{equation}
The terms with $n\ge 1$ in eq.~(\ref{Al}) yield the $O(a^{2n})$ lattice
artifacts for $m(L)L$.  As it is known, the tree level result is $A_0(L)=1$
for {\em arbitrary} lattice action --- no lattice artifacts appear in this
quantity. (It is easy to see why.  The propagator for $p=0$ states is given by
the one-dimensional propagator and $L$ enters here only as an overall factor.
Moreover, the one-dimensional lattice propagator for any regularization
differs from the continuum result $-\frac{1}{2}|t-t'|$ in terms vanishing
exponentially fast in $|t-t'|/a$. Note also that this is in accordance with
the observation that in $d=1$ any discretization of the Laplace operator is
perfect, which is related to the fact that in this case the equation $\Delta
u=0$ has only two solutions, $1$ and $t$, not infinitely many as in higher
dimensions.)

The 1-loop contribution $A_1(L)$ depends already on the form of the action 
which we write in the general form
\begin{multline} \label{AS}
{\cal A}({\bf S})= 
- \frac{1}{2}\sum_{n_1,n_2} \rho(n_1-n_2)
(1-{\bf S}_{n_1}{\bf S}_{n_2}) \\
 +\sum_{n_1,n_2,n_3,n_4} c(n_1,n_2,n_3,n_4)
(1-{\bf S}_{n_1}{\bf S}_{n_2})(1-{\bf S}_{n_3}{\bf S}_{n_4}) + \ldots
\end{multline}
In order to discuss the relation between the Symanzik tree level improved
\cite{SY}
and the FP action we shall derive first the tree level on-shell
Symanzik conditions for the action in eq.~(\ref{AS}).The $O(a^2)$ tree level
on-shell Symanzik improvement requires that all $O(a^2)$ artifacts cancel 
in all
physical quantities calculated on the tree level.
These include the spectrum $E(p)$ related to the 2-point function 
(influenced only by the coefficients $\rho$), and also the on-shell 
scattering amplitude related to the 4-point function (to which
both the coefficients $\rho$ and $c$ contribute).

An alternative way to obtain the tree level on-shell $O(a^2)$
Symanzik conditions is the following. 
Consider a lattice action ${\cal A}({\bf S})$ 
on smooth configurations ${\bf S}$ satisfying the equations of
motion, and expand it in powers of $a^2$. The coefficients of this
expansion are related to higher dimensional operators.
The $O(a^2)$ tree level Symanzik improvement is achieved 
by choosing the coefficients in ${\cal A}({\bf S})$ so that all 
the $O(a^2)$ corrections turn to zero, i.e. the lattice action 
(on the solutions ${\bf S}$)
coincides with the continuum action to this accuracy.
(For the case of SU(N) gauge theory Garcia Perez, Snippe and van Baal
\cite{Baal} have derived this expansion for a set of loops. 
From this expression one recovers the tree level result by 
L\"uscher and Weisz \cite{LW} obtained by considering 
the 2- and 3-point functions of gauge fields.)
Note that this procedure is closely related to a remarkable
property of the FP actions. Any solution of the FP lattice equations 
of motion generate a solution of the continuum theory, with exactly
the same value of the lattice and continuum actions. On the other
hand, all physical quantities calculated at the tree level using FP
actions are free of lattice artifacts. Therefore they satisfy 
the tree level Symanzik conditions to all orders $O(a^n)$
by construction.

Let us use this alternative procedure to determine the $O(a^2)$
tree level on-shell Symanzik conditions for the $O(N)$ sigma model.
By expanding the lattice derivatives one obtains
\begin{multline}\label{ASexp}
{\cal A}({\bf S}) = \frac{1}{2}\int d^2x 
(\partial_\mu {\bf S} \cdot \partial_\mu {\bf S}) + \\
 a^2 \int d^2x \left\{ 
R_1 \frac{1}{16}(\partial^2 {\bf S}\cdot\partial^2 {\bf S}) +
R_2\frac{1}{48}\sum_\mu ({\bf S}\cdot\partial_\mu^4 {\bf S}) 
\right. +   \\
 \left. C_1 \frac{1}{4}({\bf S}\cdot\partial^2 {\bf S})^2
  + C_2 \frac{1}{2} \sum_{\mu,\nu}
   (\partial_\mu {\bf S} \cdot \partial_\nu {\bf S})^2 +
C_3\frac{1}{4}\sum_\mu (\partial_\mu {\bf S} \cdot \partial_\mu {\bf S})^2
\right\} +O(a^4)\;.
\end{multline}
Here we introduced the quartic moments:
\begin{equation}
\sum_n \rho(n)n_\mu n_\nu n_\alpha n_\beta =
R_1 (\delta_{\mu\nu}\delta_{\alpha\beta}+
\delta_{\mu\alpha}\delta_{\nu\beta}+
\delta_{\mu\beta}\delta_{\nu\alpha})+
R_2 \delta_{\mu\nu\alpha\beta}\;,
\end{equation}
and
\begin{multline}
\frac{1}{V}\sum_{n_1 n_2 n_3 n_4} c(n_1,n_2,n_3,n_4)
\Delta_\mu \Delta_\nu \Delta_\alpha' \Delta_\beta' = \\
C_1 \delta_{\mu\nu}\delta_{\alpha\beta} +
C_2 (\delta_{\mu\alpha}\delta_{\nu\beta}+
\delta_{\mu\beta}\delta_{\nu\alpha})+
C_3 \delta_{\mu\nu\alpha\beta}\;,
\end{multline}
where $\Delta=n_1-n_2$, $\Delta'=n_3-n_4$, and 
$\delta_{\mu\nu\alpha\beta}$ is 1 when all its indices coincide,
otherwise zero.
Note that when restricted to solutions of the equations of motion,
$\partial^2 {\bf S} = {\bf S} ({\bf S}\cdot \partial^2 {\bf S})$,
the operators multiplying $R_1$ and $C_1$ in eq.~(\ref{ASexp}) coincide.
Accordingly, the corresponding on-shell Symanzik conditions are 
\begin{equation}
R_2=0 \;,
\label{scond1}
\end{equation}
\begin{equation}
C_1+\frac{1}{4} R_1=0,~ C_2=0,~ C_3=0 \;.
\label{scond2}
\end{equation}
These are the most general $O(a^2)$ conditions since the terms
not written out explicitly in eq.~(\ref{AS}) contribute only to 
$O(a^4)$ artifacts. Observe that the $O(a^2)$ on-shell Symanzik
conditions say nothing about $C_1$ and $R_1$ separately, only a linear
combination of these two moments enters.
The tree level spectrum is determined by the quadratic coefficients which 
has an expansion in momentum space:
\begin{equation}
\tilde{\rho}(q)=q^2 + \frac{1}{8} R_1 (q^2)^2 +
\frac{1}{24} R_2 \sum_\mu q_\mu^4 + O(q^6) \;.
\label{rtil}
\end{equation}
The absence of the $O(a^2)$ artifacts in the tree level spectrum
requires only $R_2=0$, the $(q^2)^2$ term is allowed. In this case, of course,
four-spin interactions should also be present in the action with a quartic
coupling $c$ whose moments satisfy eqs.~(\ref{scond2}). This is the
generic realization of Symanzik tree level $O(a^2)$ on-shell improvement. 
The FP action which produces no $O(a^{2n})$
artifacts satisfies the Symanzik conditions this way. On the other hand, if
one wants to have an improved action containing only two-spin interactions
then  $C_1=C_2=C_3=0$ and then eq.~(\ref{scond2}) gives $R_1=0$. In this case,
all the terms quartic in $q$ disappear in eq.~(\ref{rtil}).

We show now that the tree level on-shell Symanzik conditions cancel 
the $O(a^2)$ 
1-loop artifacts in the mass gap $m(L)$. The 1-loop contribution $A_1(L)$
to $Lm(L)$  can be written as \cite{RV}
\begin{equation}
A_1(L)=r_1(L)+(N-2)r_2(L)+s_1(L)+(N-2)s_2(L) \;.
\label{A1p}
\end{equation}
Here the terms $r_1$, $r_2$ come from the quadratic couplings
$\rho$ in eq.~(\ref{AS})
while  $s_1$ and $s_2$ from the quartic couplings $c$.
The corresponding expressions \cite{RV} could be written in a common
form:
\begin{equation}
X(L)=\frac{1}{L}\sum_{l=0}^{L-1} \int \frac{dk_0}{2\pi}
\frac{1}{\tilde{\rho}(k_0,k_1)}
F_X(k_0,k_1) \;,
\label{XL}
\end{equation}
with $k_1=\frac{2\pi}{L} l$, $X=r_1,r_2,s_1,s_2$ and
\begin{equation}
F_{r_1}(k_0,k_1)= 1-\frac{1}{2}\frac{\partial^2}{\partial k_0^2}
\tilde{\rho}(k_0,k_1) \;,
\end{equation}
\begin{equation}
F_{r_2}(k_0,k_1)= 1 
 -\delta_{k_1 0}\frac{\tilde{\rho}(k_0,0)}{\hat{k}_0^2} \;,
\end{equation}
\begin{equation}
F_{s_1}(k_0,k_1)=
-8 \left[ \frac{\partial}{\partial q_0}\frac{\partial}{\partial q_0'}
\tilde{c}(q_0,0;q_0',0;k_0,k_1) \right]_{q_0=q_0'=k_0/2}\;,
\end{equation}
\begin{equation}
F_{s_2}(k_0,k_1)=
2 \left[ 
\frac{\partial^2}{\partial q_0^2}\tilde{c}(q_0,0;0;0)-
\frac{\partial^2}{\partial q_0^2}\tilde{c}(q_0,0;k_0,k_1;0)
\right]_{q_0=0}\;.
\end{equation}
Here $\hat{k}_0^2=4\sin^2(k_0/2)$ and the Fourier transforms are
defined as
\begin{equation}
\tilde{\rho}(p)=\sum_{n}\rho(n)
{\rm e}^{-ip\cdot n}\;,
\end{equation}
\begin{equation}
\tilde{c}(p,q,r)= 
\frac{1}{V}\sum_{n_1 n_2 n_3 n_4}
c(n_1,n_2,n_3,n_4){\rm e}^{-ip\cdot \Delta}{\rm e}^{-iq\cdot \Delta'}
{\rm e}^{-ir\cdot \Delta''}\;,
\end{equation}
with $\Delta=n_1-n_2$, $\Delta'=n_3-n_4$ and
$\Delta''=(n_1+n_2-n_3-n_4)/2$.
The omitted terms in eq.~(\ref{AS})
contain at least three factors of type
$(1-{\bf S}_{n}{\bf S}_{n'})$ and hence do not contribute to $A_1(L)$.
One can explicitly
separate the leading cut-off effects in eq.~(\ref{XL}) by integrating over
$k_0$ and observing that the cut-off effects in the sum over $k_1$ are
produced by contribution from the pole at $k_0=ik_1 +\ldots$.  The final
result for the $O(a^2)$ artifacts is given by
\begin{equation}
\frac{1}{L}\sum_{l=0}^{L-1} \int \frac{dk_0}{2\pi}
\frac{1}{\tilde{\rho}(k_0,k_1)}
F(k_0,k_1) = {\rm const} -\frac{\pi}{6}\frac{\alpha}{L^2}+
O\left( \frac{1}{L^4}\right)\;,
\end{equation}
where
\begin{equation}
F(ik_1,k_1)=\alpha k_1^2 + O(k_1^4)\;,
\end{equation}
and it is also assumed that the two-point function 
$\tilde{\rho}(k)$ is improved, i.e. $R_2=0$.
After a straightforward calculation one obtains (restoring $a$)
\begin{equation}
r_1(L)+s_1(L)={\rm const} - 
\frac{\pi}{3}\left(\frac{1}{4}R_1+C_1+2C_2+C_3\right)\frac{a^2}{L^2} 
+ O\left( \frac{a^4}{L^4} \right) \;,
\end{equation}
\begin{equation}
r_2(L)+s_2(L)={\rm const} +\frac{1}{2\pi}\ln\frac{L}{a}
- \frac{\pi}{6}(2C_2 + C_3 )\frac{a^2}{L^2} 
+ O\left( \frac{a^4}{L^4} \right) \;.
\end{equation}
These equations generalize the results obtained by Caracciolo and Pelisetto
\cite{CP} for two-spin interactions.
From the general conditions 
(\ref{scond1},\ref{scond2})
it follows that there are no $O(a^2)$ artifacts at the 1-loop level
independently, whether they are realized on-shell, or off-shell.
This is consistent with the observation in 
ref.~\cite{CP} that the $O(a^2)$ artifacts in $A_1(L)$ can be cancelled by
an on-shell improved $\rho$ and an appropriately chosen quartic 
coupling $c$.
In fact, it is expected on general grounds that no physical
distinction could be made between on-shell and off-shell
improved actions. Indeed, by changing infinitesimally the field
variables in an off-shell improved action,
the resulting action will be on-shell improved since new terms
proportional to the equations of motion appear.

Let us discuss finally the conjecture in \cite{RV} that an $O(a^{2n})$
Symanzik improved action produces no cut-off effects up to $O(a^{2n})$
in the 1-loop $A_1(L)$. This suggestion has been checked in ref.~\cite{RV} 
on a specific example. We provide now a
simple argument showing that the FP action does not produce
any artifacts in $A_1(L)$.  Assume that we have calculated the mass gap to
1-loop order with the action $\beta_0 {\cal A}^{\rm FP}({\bf S})$.
(We use the notation $\beta=1/g^2$)
Alternatively, we can make first a RG step:
\begin{equation}
\beta_0 {\cal A}^{\rm FP}({\bf S}) \underset{ \text{RG}}{\longrightarrow}
\beta {\cal A}^{\rm FP}({\bf S}) + \delta{\cal A}({\bf S})
\;.
\end{equation}
The quantum corrections shift the overall coupling, $\beta_0 \to
\beta=\beta_0-\Delta\beta$, where $\Delta\beta=(N-2)\ln 2/(2\pi)$ and produce, 
in general, an extra piece $\delta{\cal A}({\bf S})$.
Calculating the mass gap with this new action one should recover the old
result including the artifacts. These are, however, associated now with the
coarser lattice, $a'=2a$. The part $\delta{\cal A}({\bf S})$ which is not
multiplied by $\beta$, should be included at tree level. Since at tree level
the artifacts to $m(L)L$ are absent for {\em any} lattice action, the quantity
$A_1(L)$ in eq.~(\ref{Al}) evaluated with ${\cal A}^{\rm FP}({\bf S})$ for
lattice spacing $a$ and $a'=2a$ should give the same artifacts.  As a
consequence, all terms in eq.~(\ref{Al}) with $n\ge 1$ should vanish, i.e.
$A_1(L)$ has no artifacts at all.  Of course, this argument does not apply to
other physical quantities having cut-off effects already on the tree level
since in this case the unknown term $\delta{\cal A}({\bf S})$ could also
contribute to the cut-off effects in the given order in $1/\beta$.

\section{The energy $E(L,p)$ in 1-loop \\
   perturbation theory }
\setcounter{equation}{0}

The amplitude $B$ and the energy $E(L,p)$ in eq.~(\ref{CC}) can be written as
\begin{equation}
B=g^2 B_0 + g^4 B_1 +\ldots ,
\end{equation}
\begin{equation}
E(L,p)=E_0 + g^2 E_1 + \ldots ,
\end{equation}
which leads to the perturbative expansion of the correlator $C(\tau ;p)$:
\begin{equation}
C(\tau ;p)= g^2 C_0(\tau ;p)+g^4 C_1(\tau ;p)+\ldots, 
\end{equation}
where, for large $\tau$ we have:
\begin{equation}
C_0(\tau ;p)= B_0 {\rm e}^{-2\tau E_0} , 
\end{equation}
\begin{equation}
C_1(\tau ;p)= (B_1-2\tau B_0 E_1) {\rm e}^{-2\tau E_0}.
\label{C1} 
\end{equation}
We shall consider $p=2 \pi/L$. The tree level propagator $C_0(\tau;p)$
defines $B_0$ and $E_0$, which can be used to determine $E_1$ from the
$\propto \tau$ part of the 1-loop two-point function $C_1(\tau;p)$.

We shall study the cut-off dependence of $E(L,p)$ using different FP actions,
Symanzik tree level improved action with next-to-nearest-neighbour (nnn)
coupling and the standard action. The FP actions give the exact 
continuum value for the tree level result $L E_0(L,p) =2\pi$, 
while the standard and the Symanzik actions have $O(a^2/L^2)$ and 
$O(a^4/L^4)$ cut-off corrections, respectively.
For the 1-loop correction of the energy $E_1$ we write
\begin{equation}
L E_1= c_0 + \frac{a^2}{L^2} c_1 + \frac{a^4}{L^4} c_2 + \ldots,
\label{LE1}
\end{equation}
where $c_0$ is the universal continuum value (turns out to be 0.5 for
$p=2\pi/L$), while $c_1,c_2,\ldots$ are numbers which depend on the form of
the action. Since $E_1$ is the leading $O(g^2)$ correction to the energy,
there are no $\ln (a^2/L^2)$ type of terms in eq.~(\ref{LE1}). The 
Feynman graph expansion for the propagator in eq.~(\ref{CC}) is the same
as for $p=0$ in the calculation of the mass gap $m(L)$. In order to 
eliminate the quasi-zero modes (they are present even at $p\neq 0$ on
the 1-loop level) we used free boundary conditions in the time direction 
\cite{LWW}. The term linear in $\tau$ in eq.~(\ref{C1}) can be separated 
analytically which leads to an easy numerical calculation. We shall summarize 
the results in Section 7.

\section{The FP action versus the marginal operator}
\setcounter{equation}{0}

In most of the applications of the FP action until now such block 
transformations were used where
the average of the fine variables in a block was allowed to fluctuate around
the block variable of the coarse lattice. This fluctuation is
governed by a parameter $\kappa$ which is optimized to obtain a short range FP
action. In the $O(N)$ $\sigma$-model the FP action satisfies the saddle-point
equation \cite{HN1}:
\begin{equation}
{\cal A}_\kappa^{\rm FP}({\bf R})= \min_{\bf S} \left[
{\cal A}_\kappa^{\rm FP}({\bf S})+ \kappa T({\bf R},{\bf S}) \right] \;,
\label{AFPK}
\end{equation}
where ${\bf R}$ and ${\bf S}$ live on the coarse and on the fine lattice,
respectively, while $T$ defines the averaging procedure.

Let us add the operator $\epsilon {\cal O}({\bf S})$ ($\epsilon$ is small) 
to the FP action and perform a RG step. 
Denoting the minimizing configuration in
eq.~(\ref{AFPK}) by ${\bf S}({\bf R})$ we get in the saddle-point 
approximation and
in linear order in $\epsilon$:
\begin{multline}
{\cal A}_\kappa^{\rm FP}({\bf S})+\epsilon {\cal O}({\bf S})
\underset{ \text{RG}}{\longrightarrow}
{\cal A}_\kappa^{\rm FP}({\bf S}({\bf R}))+
\kappa T({\bf R},{\bf S}({\bf R}))+\epsilon {\cal O}({\bf S}({\bf R})) \\
= {\cal A}_\kappa^{\rm FP}({\bf R})+\epsilon {\cal O}({\bf S}({\bf R}))
\label{EPS}
\end{multline}
Using this equation it is easy to see that ${\cal O}={\cal A}^{\rm FP}$ is 
the marginal operator of the RG transformation if $\kappa = \infty$. 
Indeed, in this limit the blocking function becomes a $\delta$-function 
constraint and we get
\begin{equation}
{\cal A}_{\infty}^{\rm FP}({\bf R})= \min_{\bf S} \left.
{\cal A}_{\infty}^{\rm FP}({\bf S}) \right|_{T({\bf R},{\bf S})=0} 
={\cal A}_{\infty}^{\rm FP}({\bf S}({\bf R})) \;.
\label{AFPI}
\end{equation}
Consequently, for ${\cal O} = A^{\rm FP}_{\infty}$ the r.h.s. 
of eq.~(\ref{EPS}) reads:
\begin{equation}
{\cal A}_{\infty}^{\rm FP}({\bf R}) +
 \epsilon{\cal A}_{\infty}^{\rm FP}({\bf S}({\bf R})) =
{\cal A}_{\infty}^{\rm FP}({\bf R}) +
 \epsilon{\cal A}_{\infty}^{\rm FP}({\bf R}) \;,
\label{AFPI2}
\end{equation}
which has the same form as the l.h.s. of eq.~(\ref{EPS}).

For finite $\kappa$ the marginal operator is not equal to the FP action,
however. In the quadratic approximation (when only the first term in
eq.~(\ref{AS}) is kept) one can construct the marginal operator explicitly for
any $\kappa$. In Fourier space the marginal operator can be written as
\begin{equation}
\rho_\kappa^{\rm marginal}(q)=
\frac{ \left( \rho_\kappa^{\rm FP}(q)\right)^2}
{\rho_{\infty}^{\rm FP}(q)}
\label{MARG}
\end{equation}

For $\kappa = \infty$ the marginal operator goes over to the FP action as
discussed in the general case above.

The formal manipulations leading to the statement that the FP action is 1-loop
perfect assume implicitly that the FP action is identical to the marginal 
operator
\cite{DHHN1}. For this reason we studied the cut-off effects in the
predictions of the FP actions in the limit $\kappa = \infty$ also. 
Using a $2\times 2$ block with a flat averaging defines a FP action in 
this limit which is rather broad \cite{HN1}. This feature does not create 
a problem when investigating the free energy density as the function of 
the chemical potential since in this calculation the quadratic couplings 
enter only. 
In the case of the energy $E(L,p)$, where the quartic couplings are needed 
to a high precision also, we considered the FP action treated in \cite{HB}. 
This action corresponds to a $\kappa = \infty$ transformation and has a short 
range $\rho$ in the notation of eq.~(\ref{AS}). This case is relevant not 
only for the issue of 1-loop perfection, but also to understand 
why the attempt to follow the RT in ref.~\cite{HB} was not successful.
The results discussed in Section 7 shed some light on this problem as well.

\section{The iterated RG transformation and \\
         the RT in 1-loop perturbation theory}
\setcounter{equation}{0}

Consider the FP action at some very large coupling $\bar{\beta}$
and perform $r$ consecutive scale=2 RG steps. (The coupling is defined, as
usual, as the coefficient of 
$\frac{1}{2}q^2 {\bf S}(q) {\bf S}(-q)$ in Fourier space.) 
Denote the finest field at the start by ${\bf S}^{(r)}$, the field 
after 1 RG step by ${\bf S}^{(r-1)}$, and the coarsest field at 
the end by ${\bf S}^{(0)}$. In the first step we obtain:         
\begin{multline}
\int D{\bf S}^{(r)} 
\exp\left\{
-\bar{\beta}\left[ {\cal A}^{\rm FP} \left( {\bf S}^{(r)}\right)
+\kappa T\left( {\bf S}^{(r-1)},{\bf S}^{(r)}\right) \right] \right\} \\
=\exp\left\{ 
-\left[ \bar{\beta}{\cal A}^{\rm FP} \left( {\bf S}^{(r-1)}\right)
+{\cal A}^{\rm q} \left( {\bf S}^{(r-1)}\right) \right] +O(1/\beta)
\right\} \;.
\label{RT1}
\end{multline}
The first term in the exponent on the r.h.s. is coming from the leading
saddle-point approximation: the contribution of the configuration
${\bf S}^{(r)}={\bf S}^{(r)}({\bf S}^{(r-1)})$ which minimizes the 
exponent on the l.h.s. (classical result). Expanding ${\bf S}^{(r)}$ around
the saddle point solution, the leading quantum corrections are independent
of $\bar{\beta}$ and are denoted by ${\cal A^{\rm q}}$ in eq.~(\ref{RT1}).

Performing the second step of RG transformation, 
$\bar{\beta}{\cal A^{\rm FP}}({\bf S}^{(r-1)})$ generates
$\bar{\beta}{\cal A^{\rm FP}}({\bf S}^{(r-2)}) + 
{\cal A^{\rm q}}({\bf S}^{(r-2)})$ as before, while 
${\cal A^{\rm q}}({\bf S}^{(r-1)})$ contributes in the leading saddle-point
approximation only giving ${\cal A^{\rm q}}({\bf S}^{(r-1)}({\bf S}^{(r-2)}))$ 
where ${\bf S}^{(r-1)}({\bf S}^{(r-2)})$  is the minimizing solution of
the saddle-point equation in the second step. After $r$ steps we get
\begin{equation}
\bar{\beta}{\cal A}^{\rm FP} \left( {\bf S}^{(r)}\right)
@>>{ r \text{ steps}}>
\bar{\beta}{\cal A}^{\rm FP} \left( {\bf S}^{(0)}\right)
+{\cal A}^{\rm q}_r \left( {\bf S}^{(0)}\right) +O(1/\beta) \;,
\end{equation}
where
\begin{equation}
{\cal A}^{\rm q}_r= {\cal A}^{\rm q}({\bf S}^{(0)})+ 
{\cal A}^{\rm q}({\bf S}^{(1)}({\bf S}^{(0)}))+ \ldots +
{\cal A}^{\rm q}({\bf S}^{(r-1)}(\ldots 
{\bf S}^{(1)}({\bf S}^{(0)})\ldots )) \;.
\label{RT3}
\end{equation}
Define $\beta = \bar{\beta} - r \Delta\beta$ with 
$\Delta\beta=\frac {N-2}{2\pi} \ln 2$ and write
\begin{equation}
{\cal A}^{\rm q}_r = -r \Delta\beta{\cal A}^{\rm FP} + 
\delta{\cal A}^{\rm q}_r \;,
\label{RT4}
\end{equation}
giving
\begin{equation}
\bar{\beta}{\cal A}^{\rm FP}\left( {\bf S}^{(r)}\right)
@>>{ r \text{ steps}}>
\beta {\cal A}^{\rm FP}\left( {\bf S}^{(0)}\right)+
\delta{\cal A}^{\rm q}_r\left( {\bf S}^{(0)}\right)+ 
O(1/\beta)\;.
\label{RT5}
\end{equation}
In the limit $r \to \infty$, $\beta$ very large, fixed 
(i.e. $\bar{\beta}\to \infty$), the r.h.s. of eq.~(\ref{RT5}) defines the RT
(the quantum perfect action). If $\delta{\cal A}^{\rm q}_{\infty}$ is neither
zero, nor redundant (i.e. it gives a non-zero contribution to the cut-off 
dependence of a physical quantity like $E(L,p)$ on the 1-loop level) then
${\cal A}^{\rm FP}$ can not be quantum perfect. In this case the predictions 
from $\beta {\cal A}^{\rm FP}$ will have cut-off effects on the 1-loop
level which are exactly compensated by the contribution from 
$\delta{\cal A}^{\rm q}_{\infty}$. In the 1-loop calculation of $E(L,p)$ only
the quadratic part of  $\delta{\cal A}^{\rm q}_{\infty}$ enters:
\begin{equation}
\label{DAQ}
\delta{\cal A}^{\rm q}_\infty\left( {\bf S}^{(0)}\right)=
-\frac{1}{2} \sum_{n,r} \delta\rho^{\rm q}_\infty (r) 
\left(1-{\bf S}^{(0)}_{n+r}{\bf S}^{(0)}_{n}\right)
+\ldots \;.
\end{equation}
As eqs.~(\ref{RT3},\ref{RT4}) show, $\delta\rho^{\rm q}_\infty$ can be obtained from
the quadratic part of the quantum correction 
${\cal A}^{\rm q} \left( {\bf S}^{(r-1)}\right)$ in eq.~(\ref{RT1}) obtained
after 1 step of RG transformation. Let us discuss briefly the steps of
this calculation.

In order to perform the path integral in eq.~(\ref{RT1}) perturbatively, 
we write
\begin{equation}
{\bf S}^{(r)}_n=\left( \sqrt{ 1- \vec{\pi}_n^2},\vec{\pi}_n \right)\;,
~~~ 
{\bf S}^{(r-1)}_{n_B}=
\left( \sqrt{ 1- \vec{\chi}_{n_B}^2},\vec{\chi}_{n_B} \right)
\;,
\label{SPI}
\end{equation}
and expand ${\cal A}^{\rm FP}$, $T$ and the measure in the fluctuations
$\vec{\pi}$ and $\vec{\chi}$. After shifting the integration variable
$\vec{\pi}_n$ by the classical solution 
$\vec{\pi}_n^{\rm c} \left( \vec{\chi}\right)$:
\begin{equation}
\vec{\pi}_n= \vec{\pi}_n^{\rm c} \left( \vec{\chi}\right)+\vec{\xi}_n \;,
\end{equation}
one obtains terms from ${\cal A}^{\rm FP}+\kappa T$ which are independent
of $\xi$, quadratic in $\xi$, etc. The terms which are independent of $\xi$
reproduce the FP action on the coarse lattice (the classical result). 
This is the first term in the exponent on the r.h.s. of eq.~(\ref{RT1}). 
The terms quadratic in $\xi$ of the type `$\chi\chi\xi\xi$' will give 
`$\chi\chi$' type of corrections after integrating over $\xi$. 
Similar contributions will be produced by the measure and by the term 
coming from the zero mode fixing, where we can replace $\vec{\pi}_n$ by 
$\vec{\pi}_n^{\rm c} \left( \vec{\chi}\right)$. (These terms are not
multiplied by $\bar{\beta}$.) Collecting all these contributions one obtains
$\rho^{\rm q}$ in ${\cal A}^{\rm q}$:
\begin{equation}
{\cal A}^{\rm q}\left( {\bf S}^{(r-1)}\right) =
-\frac{1}{2} \sum_{n_B,r_B} 
\rho^{\rm q}(r_B)\vec{\chi}_{n_B}\vec{\chi}_{n_B+r_B} +\ldots \;.
\end{equation}
The quadratic part of the quantum correction obtained after 1 step of RG
determines the quadratic part of ${\cal A}^{\rm q}_{\rm r}$ in 
eq.~(\ref{RT3}) using the relation between $\vec{\pi}_n$
and $\vec{\chi}_{n_B}$ (and its iterations) obtained from the saddle-point
equation in linear order
\begin{equation}
\vec{\pi}_n^{\rm c}=\sum_{n_B} Z\left( n-2n_B\right) \vec{\chi}_{n_B} \;,
\end{equation}
Finally, the relation eq.~(\ref{RT4}) gives $\delta\rho^{\rm q}_r$.

The RT defines a quantum perfect action which will be
used as a consistency check on the results in Section 7: the cut off
effects generated by the FP action in $E(L,p)$ should be exactly cancelled
by $\delta\rho^{\rm q}_{\infty}$. A deviation from this condition reflects
the systematic error of the calculation which is mainly due to the error
in calculating the quartic couplings $c$ in eq.~(\ref{AS}).

Let us add a remark on the finite size effects concerning the action itself. 
As discussed in the Appendix of ref.~\cite{DHHN2}, the FP action in a small
volume can be obtained from the FP action in a large volume by a simple
'wrapping' procedure. This is not true, however, for the quantum corrections,
like $\delta\rho^{\rm q}_\infty$. This correction should be calculated
separately for each small volume values. The finite size effects in the action
go to zero exponentially as the volume is increased.

\section{The free energy density as the function of the chemical potential
         in 1-loop perturbation theory with the FP action}
\setcounter{equation}{0}

The chemical potential is a very convenient tool to probe the system 
\cite{HNM,HN1}. Unlike the magnetic field, it does not get renormalized and to
obtain the free energy on the 1-loop level it is sufficient to expand the
action up to quadratic order in the fluctuations.

Technically, the chemical potential $h$ is a constant, imaginary vector
potential: $A_\mu \rightarrow ih\delta_{\mu,0}Q$, where $Q$ is the generator
of an $O(N)$ rotation. Choosing the rotation in the 0-1 plane (the $O(N)$
indices run as $i=0,1,\dots,N-1$) the effect of a non-zero chemical potential
is 
\begin{eqnarray}
S_n^0 & \to & \cosh(n_0 h) S_n^0 + i \sinh(n_0 h) S_n^1 \;,\nonumber \\
S_n^1 & \to & -i\sinh(n_0 h) S_n^0 + \cosh(n_0 h) S_n^1 \;, \label{SH} \\
S_n^j & \to & S_n^j, \qquad \text{for~} j=2,\ldots,N-1 \nonumber \;.
\end{eqnarray}
Here and in the following we use quantities whose dimension is carried by the
lattice unit. Eq.~(\ref{SH}) gives the dependence on the chemical potential of
lattice regularized actions \cite{KOG,HK}.

Consider the generalized action in eq.~(\ref{AS}). Using eq.~(\ref{SH}) and
writing ${\bf S}_n$ in terms of the fluctuations ${\bf \pi}_n$ as in
eq.~(\ref{SPI}) we get
\begin{multline}
\label{ASH}
\beta{\cal A}({\bf S}) \to Vf_0(h) + \\
\beta\frac{1}{2} \sum_{n,r}\left[ 
\rho^{(1)}(r;h)\pi_n^1\pi_{n+r}^1+
\rho^{({\rm tr})}(r;h)\sum_{j=2}^{N-1}\pi_n^j\pi_{n+r}^j
\right] + O(\pi^4)\;,
\end{multline}
where $V$ is the volume of the system and $\rho^{(1)}$ and 
$\rho^{({\rm tr})}$ are the two different $h$-dependent quadratic couplings 
($h$ breaks the symmetry between the 0-1 plane and the transversal directions).
Our task is to find $f_0(h)$, $\rho^{(1)}$ and $\rho^{({\rm tr})}$ for 
the FP action. The basic factor
$(1-{\bf S}_n {\bf S}_{n'})$ in eq.~(\ref{AS}) can be written as
\begin{equation}
\label{1SS}
1-{\bf S}_n {\bf S}_{n'} \to \left[ 1-\cosh( (n_0-n_0')h)\right]
+O(\pi^2) \;,
\end{equation}
indicating a technical problem: unlike in the case with $h=0$, the terms up to
quadratic order in the fluctuations ${\bf \pi}$ receive contributions from
terms of the action containing arbitrary many factors of the type
$(1-{\bf S}_n {\bf S}_{n'})$.

A way to proceed is to consider first an imaginary $h$
\begin{equation}
\label{HMU}
h = -i\mu
\end{equation}
which connects the problem of the free energy  
in the presence of a chemical potential with
that of the free energy of a uniformly twisted solution with fluctuations. 
Indeed, for real values of $\mu$ eq.~(\ref{SH}) at $\vec{\pi}_n=0$ describes 
such a solution
where $\mu$ is the angle (in the 0-1 plane of internal indices) 
between two neighbouring spins in the time direction,
while the configuration is constant in the spatial direction. 
This is an exact solution of the FP lattice equations of motion and 
the FP action gives the exact classical value for the action: 
$V\mu^2/2$ \cite{HN1} leading to the well-known continuum result
\begin{equation}
\label{VF0}
Vf_0(h)=-V \frac{h^2}{2 g^2} \;.
\end{equation}
For a uniformly twisted solution ${\bf R}$ with angle $2\mu$ on 
the coarse lattice, the minimizing configuration ${\bf S}$ 
in eq.~(\ref{AFPK}) is also a uniformly twisted solution with 
angle $\mu$ \cite{HN1}. 
Consider now eq.~(\ref{AFPK}) with a configuration ${\bf R}$
which contains small fluctuations around the above solution. 
Under a RG step we get in the saddle point approximation:
\begin{multline}
\label{RHOMU}
V\frac{\mu^2}{2g^2} + \frac{1}{2g^2}\sum_{n,r} 
\left[
\rho^{(1)}(r;-i\mu)\pi_n^1\pi_{n+r}^1+
\rho^{({\rm tr})}(r;-i\mu)\sum_{j=2}^{N-1}\pi_n^j\pi_{n+r}^j
\right] + O(\pi^4)  \\
\underset{ \text{RG}}{\longrightarrow}
V_B\frac{(2\mu)^2}{2g^2} + \frac{1}{2g^2}\sum_{n_B,r_B} 
\left[
\rho^{(1)}(r_B;-i2\mu)\chi_{n_B}^1\chi_{n_B+r_B}^1+ \right. \\
\left.
\rho^{({\rm tr})}(r_B;-i2\mu)\sum_{j=2}^{N-1}\chi_{n_B}^j\chi_{n_B+r_B}^j
\right] + O(\chi^4) \;,
\end{multline}
where $V_B=V/4$ is the volume of the coarse lattice.

Consider the RG transformation with the kernel
\begin{equation}
\label{BL}
\kappa T({\bf R},{\bf S})=2\kappa\sum_{n_B}
\left( {\bf R}_{n_B} -
\frac{{\bf \Sigma}_{n_B}}{|{\bf \Sigma}_{n_B}|}\right)^2 \;,
\end{equation}
where ${\bf \Sigma}_{n_B}$ is the sum over the four ${\bf S}$ spins 
in the block $n_B$. 
This kernel is a slightly modified version of the transformation used in
\cite{HN1} with the technical advantage of having a trivial norm. 
For the case of small fluctuations around the twisted solution we have
\begin{multline}
\label{KTMU}
\kappa T({\bf R},{\bf S}) \to 2\kappa \sum_{n_B} \left[
\left(\chi_{n_B}^1 -\frac{1}{4}\sum_{n\in n_B}\pi_n^1\right)^2+ \right. \\
\left. \sum_{j=2}^{N-1} \left(\chi_{n_B}^j -\frac{1}{\cos(\mu/2)}
\frac{1}{4}\sum_{n\in n_B}\pi_n^j\right)^2\right] +
(\text{quartic in the fields}) \;.
\end{multline}
It is not difficult to find the couplings $\rho^{(1)}$ and $\rho^{({\rm tr})}$ 
satisfying
eq.~(\ref{RHOMU}) under the block transformation in eq.~(\ref{KTMU}). After
continuing back to real $h$, in Fourier space they read
\begin{equation}
\label{RHO1}
\frac{1}{\rho^{(1)}(q;h)}=
\sum_{l=-\infty}^{\infty}\frac{1}{(q+2\pi l)^2}
\prod_{i=0}^1 \frac{\sin^2 \left( \frac{1}{2} q_i \right)}
{ \left( \frac{1}{2}q_i + \pi l_i\right)^2}
+\frac{1}{3\kappa} \;,
\end{equation}
\begin{equation}
\label{RHOTR}
\frac{1}{\rho^{({\rm tr})}(q;h)}=
d(h)\sum_{l=-\infty}^{\infty}\frac{1}{(q+2\pi l)^2+h^2}
\prod_{i=0}^1 \frac{\sin^2 \left( \frac{1}{2} q_i \right)}
{ \left( \frac{1}{2}q_i + \pi l_i\right)^2}
+\frac{1}{3\kappa} \;,
\end{equation}
where
\begin{equation}
\label{DH}
d(h)=\prod_{j=1}^{\infty} \frac{1}{\cosh \left( h/2^{j+1} \right)} \;.
\end{equation}
(For the transformation with non-trivial norm \cite{HN1} there enters an extra
$\cos(\mu/2)$ factor in front of the $n_B$-sum in eq.~(\ref{KTMU}) and in
eqs.~(\ref{RHO1},\ref{RHOTR}) the term $1/(3\kappa)$ is replaced by two 
different $h$-dependent functions which we do not quote here explicitly.)

Having the quadratic couplings $\rho^{(1)}$ and $\rho^{({\rm tr})}$ one should
perform a gaussian integral to get the free energy as the function of the
chemical potential. The free energy density on the 1-loop level has the form
\begin{equation}
\label{FHP}
f(h)=f(0)-h^2\left[ \frac{1}{2g^2} + 
(N-2)\frac{1}{8\pi}  \ln (c h^2) \right] + O(h^4) \;,
\end{equation}
where $c$ is a constant and the $O(h^4)$ terms represent the cut-off 
effects on the 1-loop level.
We shall discuss the results in the next section.

We close this section with remarks on testing 1-loop perfection by calculating
the free energy as the function of $h$ and on the method we applied to obtain
eqs.~(\ref{RHO1},\ref{RHOTR}). We used the trick of analytic continuation to
imaginary $h$ to connect the problem to a fluctuating
uniformly twisted solution. This solution
itself is however, unstable against certain transversal fluctuations.
(A trace of this instability is the singularity in eq.~(\ref{RHOTR}) 
at real $q$ values when $h$ is imaginary.) Of course, the original problem 
with a chemical potential is stable and we do not think that this is 
a serious problem. A somewhat more delicate question is whether the free 
energy is a good quantity to consider. We refer here to the special behaviour 
of the free energy under RG transformations \cite{KWJK}. 
A useful test would be in this context to check whether at finite $h$ 
the contribution from the quantum $\delta\rho^{\rm q}_\infty$ considered 
in Section 5 compensates exactly the cut-off effects generated by the FP 
action. This test has not been done.
 
\section{The 1-loop results}
\setcounter{equation}{0}

We present first the results on the energy $E(L,p)$ with $p=2\pi/L$ for the FP
and for the quantum perfect actions which correspond to the block 
transformation in eq.~(\ref{BL}) (trivial norm) with the optimal value 
for $\kappa$ ($\kappa=2$). In this case there are no cut-off effects on 
the tree level. The 1-loop results are summarized in Table \ref{tab:1}.

\begin{table}[ht]
  \begin{center}
    \leavevmode
    \setlength{\extrarowheight}{4pt}
    \begin{tabular}{|>{$}r<{$}|>{$}c<{$}|>{$}c<{$}|>{$}c<{$}|>{$}c<{$}|}
    \hline
      L & L E_1^{\rm FP} & (L E_1^{\rm FP}-0.5)L^2 & L E_1^{\rm q} &
                                                L(E_1^{\rm FP}+E_1^{\rm q})\\
\hline
 2 & 0.676\phantom{000} & \phantom{-}0.705\phantom{0} & -0.217\phantom{000} 
          & 0.459\phantom{000}  \\
 3 & 0.5027\phantom{00} & \phantom{-}0.0247 & \phantom{-}0.0135\phantom{00} 
          & 0.5162\phantom{00} \\
 4 & 0.494654 & -0.0855 & \phantom{-}0.006279 & 0.500933 \\
 5 & 0.496102 & -0.0975 & \phantom{-}0.004044 & 0.500146 \\
 6 & 0.497225 & -0.0999 & \phantom{-}0.002794 & 0.500019 \\ 
 7 & 0.497946 & -0.1006 & \phantom{-}0.002053 & 0.499999 \\
 8 & 0.498422 & -0.1010 & \phantom{-}0.001572 & 0.499994 \\
 9 & 0.498752 & -0.1011 & \phantom{-}0.001242 & 0.499994 \\
10 & 0.498988 & -0.1012 & \phantom{-}0.001006 & 0.499994 \\
12 & 0.499297 & -0.1013 & \phantom{-}0.000699 & 0.499996 \\
14 & 0.499483 & -0.1013 & \phantom{-}0.000513 & 0.499996 \\
\hline
\end{tabular}
\caption{
Results on the 1-loop contribution $E_1(L,p)$ to the energy $E(L,p)$
in a finite spatial volume $L$ with $p=2\pi/L$ using actions related to the
block transformation in eq.~(\ref{BL}) with $\kappa=2$. The continuum value of
$E_1(L,p)$ is 0.5, $LE^{\rm FP}_1(L,p)$ is the FP action prediction,
$LE^q_1(L,p)$ is the correction from the operator $\delta\rho^q_\infty$ in
eq.~(\ref{DAQ}) which represents the difference between the FP action and the
quantum perfect action. The deviation from 0.5 in the last column is a
systematic error due to the cuts introduced when calculating the quartic
couplings $c$ of the FP action.}
    \label{tab:1}
  \end{center}
\end{table}

The last column in Table \ref{tab:1} is the prediction of the quantum perfect
action. The deviation from the exact value (0.5) is the error of the
1-loop calculation. The source of this error is the approximation
introduced in calculating the quartic couplings of the FP action. For
$L > 6$ this error is $O(10^{-6})$ which is much smaller than the
deviation of the prediction of the FP action from 0.5 . Since for 
$L > 6$ those cut-off effects which are exponentially small (in $L$)
are negligible, we can conclude that the FP action of the block
transformation in eq.~(\ref{BL}) at $\kappa = 2$ has power like
cut-off effects in the 1-loop energy. Consequently, it is not 1-loop
perfect. The coefficient of the $O(a^2)$ cut-off correction is $-0.101$.

For comparison we give in Table \ref{tab:2} the 1-loop contribution 
$E_1(L,p)$ to the energy obtained from the standard and from the nnn 
realization of the tree level improved Symanzik actions. 
The cut-off effects seen are significantly larger than those of the FP action.
The coefficient of the $O(a^2)$ correction is 17 and 32 times larger for 
the standard and Symanzik tree level improved actions, respectively. 
It is interesting to note that cancelling the tree level $O(a^2)$ 
artifacts with an nnn interaction term in the Symanzik program
does not have any positive effect on the 1-loop artifacts -- actually, 
the cut-off effects became even larger.

\begin{table}[ht]
  \begin{center}
    \leavevmode
    \setlength{\extrarowheight}{4pt}
    \begin{tabular}{|>{$}r<{$}|>{$}c<{$}|>{$}c<{$}||>{$}c<{$}|>{$}c<{$}|}
    \hline
L & LE_1^{\rm st} & (LE_1^{\rm st}-0.5)L^2 & LE_1^{\rm SYM} &
                                       (LE_1^{\rm SYM}-0.5)L^2 \\
\hline
 2 & 0.707\phantom{000} & 0.828 & & \\
 3 & 0.6248\phantom{00} & 1.123 & & \\
 4 & 0.579786 & 1.276 & 0.143802 & -5.699 \\
 5 & 0.554845 & 1.371 & 0.349222 & -3.769 \\
 6 & 0.539873 & 1.435 & 0.406457 & -3.368 \\
 7 & 0.530223 & 1.481 & 0.434246 & -3.222 \\
 8 & 0.523654 & 1.514 & 0.450680 & -3.156 \\
 9 & 0.518991 & 1.538 & 0.461444 & -3.123 \\
10 & 0.515566 & 1.557 & 0.468957 & -3.104 \\
12 & 0.510983 & 1.582 & 0.478568 & -3.086 \\
14 & 0.508152 & 1.598 & 0.484293 & -3.079 \\
\hline
    \end{tabular}
    \caption{Results on the 1-loop contribution to the energy given 
     by the standard and by the tree level improved Symanzik actions.}
    \label{tab:2}
  \end{center}
\end{table}

As discussed in the Introduction and in Section 4, `deterministic'
($\kappa=\infty$) block transformations have the property that the marginal
operator is identical to the FP action. Since this is an assumption in the
formal considerations on 1-loop perfection \cite{DHHN1}, we tested this case
also. We considered the block transformation investigated in ref.~\cite{HB}:
the block spins sit in the even points of the fine lattice, the fine spin in
this point contributes with a weight factor of 0.8 to the block average, the
nearest-neighbours have a weight factor of 0.05. Although this transformation
is dangerously close to a decimation, it defines at $\kappa=\infty$ a FP with
a quadratic coupling $\rho$ which is short ranged \cite{HB}. Unfortunately, we
found that the quartic coupling $c$ has an extended range already and the
quantum propagator related to the fluctuations around the saddle point
solution decays less rapidly than for the block transformation in
eq.~(\ref{BL}). This is reflected by the significantly increased systematic
errors in the last column of Table \ref{tab:3} which are 
related to the cuts when calculating the quartic coupling $c$. Nevertheless,
the numbers in Table \ref{tab:3} suggest strongly that there are cut-off 
effects in the 1-loop prediction of the FP action in this case also.
\begin{table}[ht]
  \begin{center}
    \leavevmode
    \setlength{\extrarowheight}{4pt}
    \begin{tabular}{|>{$}r<{$}|>{$}c<{$}|>{$}c<{$}||>{$}c<{$}|>{$}c<{$}|}
    \hline
L & LE_1^{{\rm FP}_\infty} & (LE_1^{{\rm FP}_\infty}-0.5)L^2 
                  & LE_1^{\rm q} &  L(E_1^{{\rm FP}_\infty}+E_1^{\rm q}) \\
\hline
 2 & 1.709\phantom{000} & 2.314\phantom{0} & -0.407\phantom{000}
                                              & 0.672\phantom{000} \\
 3 & 0.5623\phantom{00} & 0.5603 & -0.0221\phantom{00} & 0.5402\phantom{00} \\
 4 & 0.52181\phantom{0} & 0.3490 & -0.01289\phantom{0} & 0.50892\phantom{0} \\
 5 & 0.512134 & 0.3034 & -0.008246 & 0.503888 \\
 6 & 0.508005 & 0.2882 & -0.005720 & 0.502285 \\
 7 & 0.505753 & 0.2819 & -0.004203 & 0.501550 \\
 8 & 0.504358 & 0.2789 & -0.003219 & 0.501139 \\
 9 & 0.503424 & 0.2773 & -0.002544 & 0.500880 \\
10 & 0.502764 & 0.2764 & -0.002060 & 0.500704 \\
12 & 0.501914 & 0.2756 & -0.001431 & 0.500483 \\
14 & 0.501404 & 0.2752 & -0.001051 & 0.500353 \\
\hline
    \end{tabular}
    \caption{Results on the 1-loop contribution to the energy using 
     the FP action and the quantum perfect action for the $\kappa=\infty$ 
     block transformation of ref.~\cite{HB}. 
     The notations are the same as in Table \ref{tab:1}.}
    \label{tab:3}
  \end{center}
\end{table}

For further clarification we studied the free energy density as the function of
the chemical potential as discussed in Section 6. From eq.~(\ref{FHP}) follows
that the combination
\begin{equation}
R(h)=\frac{d}{dh^2} \left( f(h)+\frac{h^2}{2g^2}\right) +
(N-2)\frac{1}{8\pi}\ln (h^2) \;.
\label{RH}
\end{equation}
is a constant up to cut-off effects. Table \ref{tab:4} shows the results 
using the FP action generated by the block transformation in eq.~(\ref{BL}) 
with $\kappa=\infty$. This calculation has practically no systematic
errors. (See, however, the remarks at the end of Section 6.) 
$R(h)$ shows small deviations from a constant leading to the conclusion
that even those FP actions which correspond to $\kappa=\infty$ 
(`deterministic') block transformations produce non-zero (although small) 
cut-off effects in 1-loop perturbation theory.
\begin{table}[ht]
  \begin{center}
    \leavevmode
    \setlength{\extrarowheight}{4pt}
    \begin{tabular}{|>{$}c<{$}|>{$}c<{$}||>{$}c<{$}|>{$}c<{$}|
                    |>{$}c<{$}|>{$}c<{$}|}
    \hline
    h  & R(h) & h  & R(h) & h  & R(h) \\
    \hline
1.4 & 0.142299 & 0.8 & 0.138704 & 0.4 & 0.137272 \\
1.2 & 0.140938 & 0.7 & 0.138265 & 0.3 & 0.137057 \\
1.0 & 0.139732 & 0.6 & 0.137878 & 0.2 & 0.136902 \\
0.9 & 0.139194 & 0.5 & 0.137546 & 0.1 & 0.136809 \\
\hline
    \end{tabular}
    \caption{The values of $R(h)$ from eq.~(\ref{RH}) for different
             values of the chemical potential $h$}.
    \label{tab:4}
  \end{center}
\end{table}

Acknowledgement: The authors are indebted for discussions with T.~DeGrand,
F.~Farchioni, A.~Hasenfratz, V.~Laliena, A.~Papa and P.~Weisz. 
A part of this work was done while one of us (P.H.) visited the Departament
d'Estructura i Constituents de la Materia, University of Barcelona in
the program `Professores Visitantes' supported by Iberdrola S.A.. He
thanks for the warm hospitality and for the discussions with D.~Espriu,
J.I.~Latorre and A.~Travesset during this time.
This work has been completed while the other author (F.N.) enjoyed
the hospitality of the Newton Institute, Cambridge. 
He thanks P.~van~Baal for the discussions there.


\end{document}